\documentclass[10pt]{article}
\usepackage[margin=1in]{geometry}
\usepackage{amsmath}
\usepackage{setspace}
\usepackage{graphicx}
\usepackage{xcolor,colortbl}
\usepackage{tabu}
\usepackage{tikz}
\usepackage{abstract}
\usetikzlibrary{trees}
\definecolor{light-gray}{gray}{0.95}
\title{Bounds on the Number of Huffman \\ and Binary-Ternary Trees}
\author{Angeline Rao[1], Ying Liu[2], Yezhou Feng[3], and Jian Shen[4]\\ \\
Clements High School, Sugar Land, Texas 77479[1], \\
Liberal Arts and Science Academy, Austin, Texas 78724[2],\\ 
Memorial High School, Houston, Texas 77024[3], \\
and Department of Mathematics, Texas State University, San Marcos, Texas 78666[4], \\
js48@txstate.edu\\}
\begin{document}
\begin{spacing}{2.0}
\maketitle
%

\begin{abstract}
Huffman coding is a widely used method for lossless data compression because it optimally stores data based on how often the characters occur in Huffman trees. An $n$-ary Huffman tree is a connected, cycle-lacking graph where each vertex can have either $n$ ``children" vertices connecting to it, or 0 children. Vertices with 0 children are called \textit{leaves}. We let $h_n(q)$ represent the total number of $n$-ary Huffman trees with $q$ leaves. In this paper, we use a recursive method to generate upper and lower bounds on $h_n(q)$ and get $h_2(q) \approx (0.1418532)(1.7941471)^q+(0.0612410)(1.2795491)^q$ for $n=2$. This matches the best results achieved by Elsholtz, Heuberger, and Prodinger in August 2011. Our approach reveals patterns in Huffman trees that we used in our analysis of the Binary-Ternary (BT) trees we created. Our research opens a completely new door in data compression by extending the study of Huffman trees to BT trees. Our study of BT trees paves the way for designing data-specific trees, minimizing possible wasted storage space from Huffman coding. We prove a recursive formula for the number of BT trees with $q$ leaves. Furthermore, we provide analysis and further proofs to reach numeric bounds. Our discoveries have broad applications in computer data compression. These results also improve graphical representations of protein sequences that facilitate in-depth genome analysis used in researching evolutionary patterns.
\end{abstract}
\end{spacing}

\begin{spacing}{2.0}

\section{Introduction}
\subsection{Background Information}
Huffman trees originated in 1952, when David A. Huffman devised an algorithm for lossless data compression that produces an optimal, prefix-free replacement code that is represented by a Huffman tree [1]. Huffman coding is ``optimal" in the sense that it minimizes the expected word length. Additionally, the coding is ``prefix-free" in that no code word forms the prefix of any other code words. Because of the significance of Huffman coding, Huffman trees and their patterns have been extensively analyzed. The enumeration of Huffman sequences is frequently studied by researchers.

\subsection{Applications}
Huffman trees have applications in a plethora of diverse fields. Most significantly, Huffman coding is a commonly used method of data compression that serves as the crucial final step in compressing MP3 files, MPEG files, and JPG images to minimize the file size. Furthermore, the Huffman algorithm is a key component of the DEFLATE algorithm, which compresses ZIP files and PNG images. Additionally, the Huffman coding method implicates advances in biochemical fields. In a method formulated in [2], a binary Huffman tree is used for coding amino acids based on their frequencies. The amino acid codes are then used to construct a lossless, 2-D graphical representation of proteins, which was never possible in the past. \\
\indent Our research facilitates further opportunities to capitalize on the efficiency and simplicity of Huffman coding by computing all possible sets of codes for a fixed number of characters. Furthermore, our research on BT trees opens a whole new door in data compression by providing a new data structure with the potential to resolve the existing setbacks of Huffman trees. Given a long alphabet, the conventional binary Huffman tree generates codes of excessive length. In higher-order Huffman trees, it is common for there to be leftover leaves that are don't represent anything, which results in wasted storage space. Our results provide an important step toward solving these issues by introducing the idea of constructing trees based on the specific type of data that is being compressed.

\section{Definitions and Previous Work}

\subsection{Definitions}

Before presenting our research, we need to define several key terms, functions, and sets.
A $n$-ary Huffman tree has one vertex on its lowest level called the \textit{root} that has $n$ children located on the next level (level 1). Every other vertex  can have either $n$ or 0 children. Vertices that don't have children are called \textit{leaves}. All other vertices have a degree of $n+1$ and are called \textit{internal vertices}.
\begin{center}
\begin{tikzpicture}[level distance=1.3cm,
  level 1/.style={sibling distance=2.3cm},
  level 2/.style={sibling distance=1.6cm},
  level 3/.style={sibling distance=1.1cm}]
  \node [circle,draw, thick, minimum width=3mm](z){}
    child {node[circle,draw, thick, minimum width=3mm]{ }
      child {node [circle,draw, thick, minimum width=3mm]{ }}
      child {node [circle,draw, thick, minimum width=3mm]{ }}
    }
    child {node[circle,draw, thick, minimum width=3mm](z1){ }
    child {node [circle,draw, thick, minimum width=3mm]{}
    child {node [circle,draw, thick, minimum width=3mm]{ }}
      child {node [circle,draw, thick, minimum width=3mm]{ }}
    }
      child {node [circle,draw, thick, minimum width=3mm](z2){ }
      child {node [circle,draw, thick, minimum width=3mm]{ }}
      child {node [circle,draw, thick, minimum width=3mm](z3){ }}
      }
    };
    \node[right=.2cm, black] at (z) {  $\rightarrow$ level 0, lowest level};
    \node[right=.2cm, black] at (z1) {  $\rightarrow$ level 1};
    \node[right=.2cm, black] at (z2) {  $\rightarrow$ level 2};
    \node[right=.2cm, black] at (z3) {  $\rightarrow$ level 3, highest level};

\end{tikzpicture}
\label{figure:hi}\\
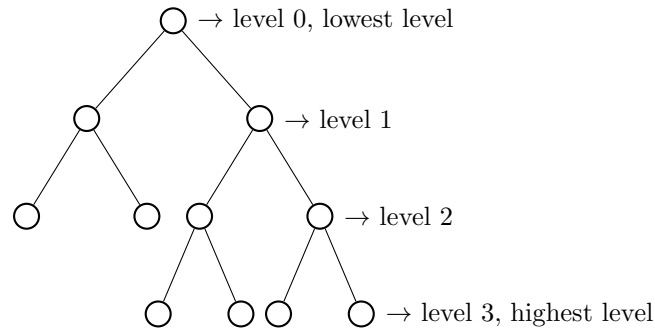
\footnotesize{Figure 2.1.1. Example of a binary Huffman tree.}
\end{center}
Figure 2.1.1 illustrates a binary Huffman tree with 6 leaves, two on level 2 and four on level 3. The single vertex on level 0 is the root.
The total number of leaves in a full $n$-ary tree is denoted by $q$. The number of internal vertices on the second highest level, or the number of parents that have children on the highest level, is denoted by $s$. For example, in Figure 2.1.1, $s$ is equal to 2. Thus $ns$ is the total number leaves on the top level, which we call $p$. A \textit{path} from one vertex to another is made up of existing edges and vertices and can only bypass each level once. The length of such a path is the number of edges that it is composed of.\\
\begin{center}
\begin{tikzpicture}
[level distance=1.3cm,
  level 1/.style={sibling distance=2.5cm},
  level 2/.style={sibling distance=1.8cm},
  level 3/.style={sibling distance=1.2cm},
  arrow/.style={edge from parent/.style={draw,-latex}}
  ,
  noarrow/.style={edge from parent/.style={draw,-}}
  ]
  \node [black,fill=blue!50,circle,draw,thick,minimum width=3mm](r){  }
    child[blue!50,arrow] {node[black,fill=blue!50,circle,draw,thick,minimum width=3mm]{ }
      child [black, noarrow]{node [circle,draw,thick,minimum width=3mm]{ }
          child{node [circle,draw,thick,minimum width=3mm]{ }}
          child{node [circle,draw,thick,minimum width=3mm]{ }}
      }
      child [blue!50,arrow]{node [black,fill=blue!50,circle,draw,thick,minimum width=3mm]{ }
          child[black, noarrow]{node [circle,draw,thick,minimum width=3mm]{ }}
          child{node [black,fill=blue!50,circle,draw,thick,minimum width=3mm](a){ }}}
    }
    child{node[circle,draw,thick,minimum width=3mm]{ }
        child {node [circle,draw,thick,minimum width=3mm]{ }}
      child {node [circle,draw,thick,minimum width=3mm](b){ }}
    };

\end{tikzpicture} \\

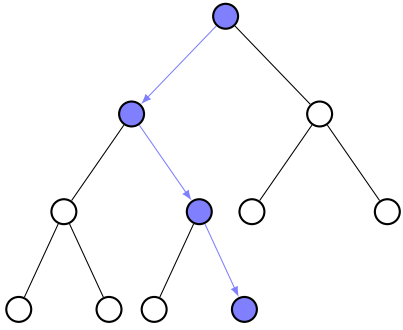
\footnotesize{Figure 2.1.2. The blue edges and vertices indicate an example of a root-to-leaf path of length 3.}
 \end{center}

An $n$-ary Huffman sequence of length $q$ is defined as a non-decreasing list of the root-to-leaf path lengths for each of the $q$ leaves in a full $n$-ary tree. Thus the Huffman sequence for Figure 2.1.2 would be \{2, 2, 3, 3, 3, 3\}. We define $h_n(q)$ to be the total number of $n$-ary Huffman sequences of length $q$.

Two trees are \textit{Huffman-equivalent} if their Huffman sequences are identical. For example, since the Huffman sequences of the trees in Figures 2.1.1 and 2.1.2 are both \{2, 2, 3, 3, 3, 3\}, we count these two trees as one. $T_n(p,q)$ is the set of Huffman-equivalence classes of full $n$-ary trees with $q$ total leaves, $p$ of which are on the top level. $t_n(p,q)$ is equal to $|T_n(p,q)|$. Thus $h_n(q)$ equivalently represents the total size of all possible equivalence classes of $n$-ary Huffman trees with $q$ leaves. This is represented as $h_n(q)=\displaystyle\sum\limits_{p \ge n} t_n(p,q)$ because the right hand side encompasses all possible values of $p$. 

We let $k$ be the number of vertices with degree $n+1$ in a full $n$-ary tree. The equation $q = n+k(n-1)$ can be easily reached by noting that the sum of the degrees of all vertices is double the number of edges, and that the number of edges is one less than the total number of vertices. $H_n(k)$ also represents the number of $n$-ary Huffman sequences.
\subsection{Previous Work}

Prior to and during our research, we closely studied some recent work done on enumerating Huffman trees. These papers helped us gain insight on discovering new patterns to improve the current enumeration techniques used for studying Huffman sequences. Furthermore, they gave us inspiration about approaches to use when taking the challenge of enumerating Binary-Ternary trees.
\subsubsection{Paschke, Bukert, and Fehribach's Work [4]}

\paragraph{Theorem 2.1.} $\displaystyle\sum\limits_{p \ge s} t_n(p,q)=t_n(ns,q+(n-1)s)$. 
\paragraph{\textbf{Proof:}} Paschke, Bukert, and Fehribach illustrated a bijection between $\displaystyle\bigcup\limits_{p \ge s} T_n(p,q)$ and $T_n(ns,q+(n-1)s)$. They showed a one-to-one correspondence by pointing out that for a tree $T$ such that $T \in \displaystyle\bigcup\limits_{p \ge s} T_n(p,q)$, $n$ children can be given to $s$ of the $p$ leaves on the highest level, forming an element of $T_n(ns,q+(n-1)s)$. For example, one of the four leaves on the highest level are chosen and given two children as shown in Figure 2.2.1.

They then showed that removing the $ns$ leaves from the highest level of $T'$ such that $T' \in T_n(ns,q+(n-1)s)$ results in an element of $\displaystyle\bigcup\limits_{p \ge s} T_n(p,q)$. In Figure 2.2.1., going in reverse, the two leaves on the top level of the tree on the right are removed and we are back to an element of the original set of trees. This inverse operation proves that the mapping between the sets is onto as well. Therefore, the sets are equal in size.

\begin{center}
\begin{tabular}[t]{p{4cm}p{1.5cm} p{5cm}}
    \vspace{0pt}   

\begin{tikzpicture}[level distance=1.3cm,
  level 1/.style={sibling distance=2cm},
  level 2/.style={sibling distance=1.5cm}]
  \node [circle,draw, thick, minimum width=3mm](z){  }
    child {node[circle,draw, thick, minimum width=3mm]{ }
      child {node [circle,draw, thick, minimum width=3mm]{ }}
      child {node [circle,draw, thick, minimum width=3mm]{ }}
    }
    child {node[circle,draw, thick, minimum width=3mm]{ }
    child {node [circle,draw, thick, minimum width=3mm]{ }}
      child {node [circle,draw, thick, minimum width=3mm]{ }}
    };
\end{tikzpicture}
\;
&

\vspace{0pt}
\hspace*{0.5cm}
\vspace{25pt}
\begin{tikzpicture}[baseline=1.5cm]
\node (a) at (-1, 0) {};
\node (b) at (1,  0) {};
\draw[-latex] (a) -- (b);
\draw[-latex] (b) -- (a);
\end{tikzpicture}
\;
&

\vspace{0pt}
\hspace*{1.5cm}
\begin{tikzpicture}[level distance=1.3cm,
  level 1/.style={sibling distance=2cm},
  level 2/.style={sibling distance=1.5cm},
  level 3/.style={sibling distance=1.5cm}
  ]
  \node [circle,draw, thick, minimum width=3mm]{  }
    child {node[circle,draw, thick, minimum width=3mm]{ }
      child {node [circle,draw, thick, minimum width=3mm]{ }
      child{node [fill=yellow,circle,draw, thick, minimum width=3mm]{ }}
      child{node [fill=yellow,circle,draw, thick, minimum width=3mm]{ }}
      }
      child {node [circle,draw, thick, minimum width=3mm]{ }
      child [missing]
      child [missing]}
    }
    child[] {node[circle,draw, thick, minimum width=3mm]{ }
        child {node [circle,draw, thick, minimum width=3mm]{ }}
      child {node [circle,draw, thick, minimum width=3mm]{ }}
    };
\end{tikzpicture}
\end{tabular} \\
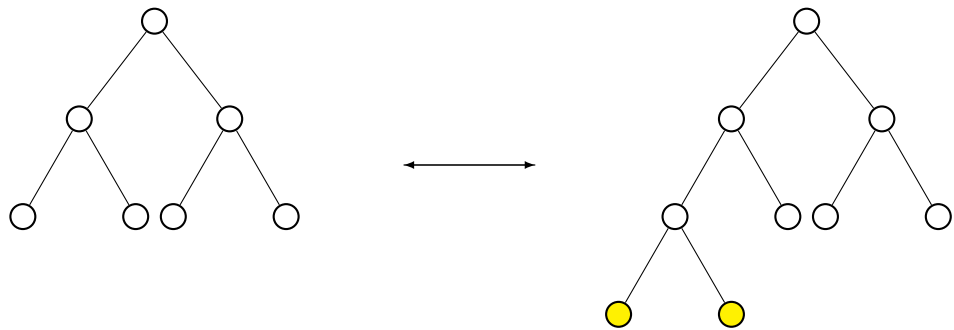
\scriptsize{Figure 2.2.1. An illustration of Theorem 2.1}

\end{center}

\paragraph{Theorem 2.2.}

\textit{If $n\ge2$ and $k\ge n+1$ then $H_n(k) \le \displaystyle\sum\limits_{i=1}^{n+1} H_n(k-i)$, and if $k \ge n+3$ then $H_n(k)\ge \displaystyle\sum\limits_{i=1}^n H_n(k-i)+H_n(k-(n+2))$.}
\\ \newline
Using applications of Theorem 2.1, the equation $q=n+k(n-1)$ and $H_n(k)$ as defined previously, Panschke, Bukert, and Fehribach were able to reach these results, which led to the inspiration of the method we use to compute upper and lower bounds on the number of  $n$-ary Huffman sequences.


\subsubsection{Elsholtz, Heuberger, and Prodinger's Work [5]}
In 2011, Elsholtz, Heuberger, and Prodinger were able to prove a very accurate asymptotic result on $h_n(q)$ by using generating functions and the asymptotic growth of $h_n(q)$. Their method is able to compute approximations for $R,R_2,c_1,c_2$ such that $h_n(q) \approx c_1R_1^q+c_2R_2^q$ for any value of $n$.\subsection{New Results}
In this paper, we provide and prove a simpler, recursive method that reaches the approximation of $h_n(q)$ in [5] and also exposes new patterns in Huffman trees. We then exploit these patterns in our analysis of Binary-Ternary (BT) trees and prove a recursive formula for the number of BT trees with $q$ leaves. Finally, we offer insight as to how numerical bounds for BT trees can be reached.

\section{Bounds on the Number of n-ary Huffman Sequences}
\subsection{Introduction}

In this section we prove a recursive method of generating upper and lower bounds for $h_n(q)$ that converge to the approximation reached by Elsholtz et al. Our recursive method reveals new patterns in Huffman trees and provides a strong base for the study of more complex trees that can better compress data.
\subsection{Proof of our Results}

\paragraph{Theorem 3.1} $t_n(ns,q)=h_n(q-s(n-1))-\displaystyle\sum\limits_{j=1}^{\lfloor \frac{s-1}{n} \rfloor} t_n(nj, q-s(n-1))$.

\paragraph{Proof:} It follows from Theorem 2.1 that $t_n(ns,q)=h_n(q-s(n-1))-\displaystyle\sum\limits_{p<s} t_n(p,q-(n-1)s)$. To figure out what to subtract from $h_n(q-s(n-1))$, we need to compare values of $s$ to possible values of $p$. Since the smallest value of $p$ is always $n$, for all cases of $s=1,2,...,n$, it is true that $t_n(p,q-(n-1)s)=h_n(q-s(n-1))$. Likewise, for the next $n$ terms, namely $s=n+1, n+2, ..., 2n$, the case where $p=n$ is going to be smaller than these cases of $s$. Thus for the second group of $s$ values, it follows that \\ $t_n(p,q-(n-1)s)=h_n(q-s(n-1))-t_n(n,q-(n-1)s)$. To generalize this pattern, we find that the general formula for every $n$ terms is the same, which suggests dividing by $n$ and taking the floor function. However, this causes problems when $s=n$ and we get $\lfloor \frac{n}{n} \rfloor =1$ when we want it to be 0. To fix this, we subtract one before we divide, and obtain the desired result.
\newline
From the results in Theorem 3.1, we can create charts where every $t_n(ns,q)$ term is represented by the sum of $h_n(q-i)$ terms.
\begin{spacing}{1.5}
\begin{center}

\begin{tabular}{rrrrrrrrrrrrrrrr}
\hline\hline                     
   $i$ &          1 &          2 &          3 &          4 &          5 &          6 &          7 &          8 &          9 &         10 &         11 &         12 &         13 &         14 &         15 \\
\hline          

 $t(2,q)$ &         \cellcolor{gray!15} 1 &        \cellcolor{gray!15}\cellcolor{gray!15} &        \cellcolor{gray!15}    &            &            &            &            &            &            &            &            &            &            &            &            \\

$t(4,q)$ &         \cellcolor{gray!15}   &      \cellcolor{gray!15}\cellcolor{gray!15}    1 &    \cellcolor{gray!15}     &            &            &            &            &            &            &            &            &            &            &            &            \\
    $t(6,q)$ &            &            &        \cellcolor{gray!40}  1 &       \cellcolor{gray!40}  -1 &      \cellcolor{gray!40}       &            &            &            &            &            &            &            &            &            &            \\

    $t(8,q)$ &            &            &          \cellcolor{gray!40}   &         \cellcolor{gray!40} 1 &        \cellcolor{gray!40} -1 &            &            &            &            &            &            &            &            &            &            \\

   $t(10,q)$ &            &            &            &            &  \cellcolor{gray!15}    1 &   \cellcolor{gray!15}    -1 &     \cellcolor{gray!15}  -1 &      \cellcolor{gray!15}      &            &            &            &            &            &            &            \\

   $t(12,q)$ &            &            &            &            &     \cellcolor{gray!15}\cellcolor{gray!15}       &    \cellcolor{gray!15}      1 &     \cellcolor{gray!15}    -1 &     \cellcolor{gray!15}    -1 &            &            &            &            &            &            &            \\

   $t(14,q)$ &            &            &            &            &            &            &        \cellcolor{gray!40}  1 &     \cellcolor{gray!40}    -1 &     \cellcolor{gray!40}    -1 &     \cellcolor{gray!40}    -1 &     \cellcolor{gray!40}     1 &  \cellcolor{gray!40}           &            &            &            \\

   $t(16,q)$ &            &            &            &            &            &            &         \cellcolor{gray!40}    &    \cellcolor{gray!40}     1 &         \cellcolor{gray!40}-1 &   \cellcolor{gray!40}      -1 &   \cellcolor{gray!40}      -1 &    \cellcolor{gray!40}      1 &            &            &            \\
 
\hline

\end{tabular}
\end{center}
\begin{center}
\footnotesize{Table 3.1. This is a table that represents $t_2(p,q)$ as $h_2(q-i)$ terms.}
\end{center}
\end{spacing}
\begin{spacing}{2.0}
            Each term in the table can be generated recursively. For example, if the representation of $t_n(6,q)$ is desired, then by Theorem 3.1 we know that $\lfloor \frac{3-1}{2} \rfloor=1$ case of $s$, namely $s=1$, must be subtracted from $h_2(q-3)$. According to the chart, $t_2(2,q)=h_2(q-1)$, so $t_2(2,q-3)=h_2(q-3-1)=h_2(q-4)$.

\subsection{The Bounds.}
Our bounds are generated based upon a given integer $i$, meaning that only the cases where $s=1,2,...,i$ in $t_n(ns,q)$ will be used in the computation of the bounds. Using such a variable greatly increases efficiency by allowing the bounds of any desired specificy to be computed. For example, if the value $h_3(15)$ is desired, there is no need to compute a recursive chart beyond the fourth row, the row representing $t_3(3(4),15)$, because the rest of the rows represent degenerate terms.\\
\indent It is important to note that there are restrictions on $i$. For example, there is no ternary tree that has 8 leaves. Here we note the common relationship between $q$, $k$, and $n$ noted in Theorem 2.2. Since $q=n+k(n-1)$ can be alternatively expressed as $q=(k+1)(n-1)+1$, it follows that $q \equiv 1$ (mod $n-1$). Since we also want $q-i \equiv 1$ (mod $n-1$) to be true, $n-1$ must divide $i$.

\subsubsection{The Lower Bound} We know that $h_n(q)=\displaystyle\sum_{s=1}^{i} t_n(ns,q)+\displaystyle\sum\limits_{s \ge i+1} t_n(ns,q)$, for some integer $i$. Since $\displaystyle\sum\limits_{s \ge i+1} t_n(ns,q)$ must be nonnegative because it counts the number of trees that satisfy a certain condition, $\displaystyle\sum\limits_{s=1}^{i} t_n(ns,q) \le h_n(q)$. To compute the lower bound, we add up the first $i$ rows in the chart to get a recursive formula for $h_n(q)$. Then, we use the coefficients to determine a characteristic equation that we solve to get the explicit formula for a lower bound.
For higher values of $q$, in order to make the computation easier, we design a program that computes a list of coefficients for a given integer $i$ where the $k$th coefficient corresponds to $h_n(q-k)$. We first create an array representing the values of $t_n(p,q)$, called $t_n(sn,q)$[] for any given $p$. Then we set $t_n(sp,q)[s]=1$, and for every $t_n(sp,q)[k]$, where $k>n$, we let $t_n(sp,q)[k] = -\displaystyle\sum\limits_{i=1}^{k-1} t_n(sp,q)[i].$ We then create another array, Cof[], such that Cof[k]=$\displaystyle\sum\limits_{i=1}^{k-1} t_n(sp,q)[i]$. This final array then represents the recursive coefficients for a lower bound on $h_n(q)$.
 
\subsubsection{The Upper Bound} Because \\ $h_n(q-i-(n-1))=t_n(n,q-i-(n-1))+\displaystyle\sum\limits_{p=2n}^{i} t_n(p,q-i-(n-1))+\displaystyle\sum\limits_{p \ge i+(n-1)} t_n(p,q-i-(n-1))$, it follows that
\begin{center}
$h_n(q-i-(n-1)) \ge t_n(n,q-i-(n-1))+\displaystyle\sum\limits_{p \ge i+(n-1)} t_n(p,q-i-(n-1))   $. \colorbox{light-gray}{(1)}
\end{center}
By the same logic, $h_n(q-i-2(n-1))\ge t_n(n,q-i-2(n-1))+\displaystyle\sum\limits_{p \ge i+2(n-1)} t_n(p,q-i-2(n-1))$. Since we know by Theorem 3.1 that $t_n(n,q-i-(n-1))=h_n(q-i-2(n-1))$, substituting this into (1) yields that
\begin{align*}
h_n(q-i-(n-1))\ge \left(t_n(n,q-i-2(n-1))+\displaystyle\sum\limits_{p \ge i+2(n-1)} t_n(p,q-i-2(n-1))\right) \\ +\displaystyle\sum\limits_{p \ge i+(n-1)} t_n(p,q-i-(n-1)).
\end{align*}
The second and third terms on the right hand side are in the summation $h_n(q)=\displaystyle\sum\limits_{s \ge 1} t_n(ns,q)$. We can continually generate equations of the form $h_n(q-k)\ge t_n(n,q-k)+\displaystyle\sum\limits_{p \ge k} t_n(p,q-k)$ for all values of $k \ge i+1$. Then, after substituting all these equations into \colorbox{light-gray}{(1)}, the result is
\begin{align*}
h_n(q-i-1) &\ge \displaystyle\sum\limits_{p \ge i+(n-1)} t_n(p,q-i-(n-1))
\\&+ \displaystyle\sum\limits_{p \ge i+2(n-1)} t_n(p,q-i-2(n-1))
\\&+ \displaystyle\sum\limits_{p \ge i+3(n-1)} t_n(p,q-i-3(n-1))
\\&+ \displaystyle\sum\limits_{p \ge i+4(n-1)} t_n(p,q-i-4(n-1))
\\&...
\end{align*}
This shows that $h_n(q-i-(n-1))$ is greater than all the terms that the lower bound neglects to account for. Thus $\displaystyle\sum\limits_{s=1}^{i} t_n(ns,q)+h(q-i-(n-1)) \ge h_n(q)$. To compute this bound, the first term in the $(i+1)$st row of the chart is added to the lower bound. Alternatively, 1 can be added to the $(i+(n-1))$th term in the list of coefficients for the lower bound. \\
The difference between the bounds is $h_n(q-i-(n-1))$. The value of $h_n(q-i-(n-1))$ will become smaller and eventually reach zero as $i$ approaches $q$, showing that the upper and lower bounds converge. Knowing this, we can reach an approximation of $h_n(q)$ by solving the characteristic equation to get the values of $r$ such that $h_n(q) \approx c_1r_1^q+c_2r_2^q+...$. To solve for the $c_k$ constants, we find the value of $h_n(q)$ for a large value of $q$ and then plug in arbitrarily large values of $i$. Realistically, however, the $c_k$ is extremely difficult to calculate beyond $k=2$, the result that Elsholtz et al. reached.

For example, when we use this method to calculate the numeric bounds for binary trees, we first solve the characteristic equation for the two largest roots. We then compute $h_2(q)$ for arbitarily large values of $q$ to determine the value of the $c_1$ and $c_2$. Our result is
\begin{center}
$h_2(q) \approx (0.1418532)(1.794147187541)^q+(0.061241041)(1.2795491341242096)^q$
\end{center}
This approximation is the same that was reached by Elsholtz et al. [5]. Similar steps can be taken for higher values of $n$. 
\section{Analysis of Binary-Ternary trees}
\subsection{Introduction} 
In order to solve some of the issues that Huffman trees pose, such as the potential wasted storage space, we decided to create a new data structure. In this type of tree, the number of children a parent can have varies from level to level. We study trees with vertices that can have either $0$, $2$, or $3$ children, depending on the level that the vertex is located on. We call these trees \textit{Binary-Ternary trees}. Our first area of study focuses on trees where the number of children each vertex can have depends on the parity of its level. In other words, vertices on even levels, including level 0, can have two or zero children, while vertices on odd level can have either three or zero children. We refer to these trees specifically as 2-3 trees, as shown in Figure 4.1.
\begin{center}
\begin{tikzpicture}[level distance=1.2cm,
  level 1/.style={sibling distance=2.9cm},
  level 2/.style={sibling distance=1.2cm},
  level 3/.style={sibling distance=0.7cm},
  level 4/.style={sibling distance=0.7cm}]
  \node [black, circle,draw, thick, minimum width=2mm](z){ }
    child {node[black, circle,draw, thick, minimum width=2mm]{ }
      child {
      node [black, circle,draw, thick, minimum width=3mm]{ }
       child{node [circle,draw, thick, minimum width=3mm]{ }}
       child{node [circle,draw, thick, minimum width=3mm]{ }}
      }
      child {node [black ,circle,draw, thick, minimum width=3mm]{ }
       child{node [circle,draw, thick, minimum width=3mm]{ }}
       child{node [circle,draw, thick, minimum width=3mm]{ }}
       }
      child {node [black, circle,draw, thick, minimum width=3mm]{ }}
    }
    child {node[black, circle,draw, thick, minimum width=3mm]{ }
    child {node [black,circle,draw, thick, minimum width=3mm]{ }}
      child {node [black,circle,draw, thick, minimum width=3mm]{ }}
      child {node [black, circle,draw, thick, minimum width=3mm]{ }}
    };
\end{tikzpicture}
\end{center}
\begin{center}
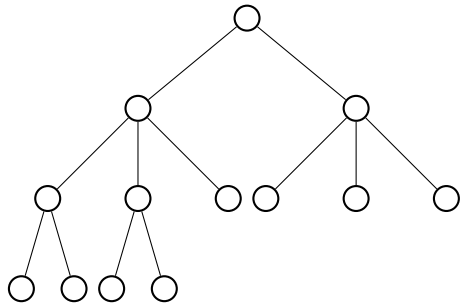
\footnotesize{Figure 4.1 This is an example of a 2,3 tree.}
\end{center}

Because the parity of the level determines how many children every vertex on that level can have, we break up the $2,3$ trees into two categories: those with an odd top level and those with an even top level. The number of $2,3$ trees with $q$ total leaves and an even or odd top level is represented by $e_{2,3}(q)$ or $o_{2,3}(q)$, respectively. The number of $2,3$ trees with $p$ out of $q$ total leaves on the even top level is denoted by $t_{2,3}(p,q_e)$, whereas if the highest level is odd, the quantity is represented by $t_{2,3}(p,q_o)$. $T_{2,3}(p,q)$ is the set of equivalence classes of $2,3$ trees that have $p$ leaves on the highest level. 
\newline \indent We are the first people to study BT trees as far as we know. In this section, we prove a recursive formula for the number of $2,3$ trees with $q$ leaves. Furthermore, we analyze and provide further proofs as to how numerical bounds of BT trees can be reached.

\subsection{The Proof}
\paragraph{Theorem 4.1.1} $t_{2,3}(2s,q_o)=(q-s)_e-\displaystyle\sum\limits_{p=1}^{\lfloor \frac{s-1}{3} \rfloor} t_{2,3}(p,(q-s)_e)$
\paragraph{Proof:} When building a $q_o$ tree, we think of it as starting out with a $(q-s)_e$ tree with $p$ leaves on its highest level, where $p \ge s$, and then giving two children to $s$ of the $p$ top-level leaves. This results in a tree with $q$ leaves and an odd top level. \\
\begin{center}
\begin{tikzpicture}[
  level distance=1.2cm,
  level 1/.style={sibling distance=2.9cm},
  level 2/.style={sibling distance=1.2cm},
  level 3/.style={sibling distance=0.7cm},
  level 4/.style={sibling distance=0.7cm}]
  \node [align=left,black, fill=blue!50,circle,draw, thick, minimum width=2mm](z){ }
    child {node[black, fill=blue!50, circle,draw, thick, minimum width=2mm]{ }
      child {
      node [black, fill=blue!50,circle,draw, thick, minimum width=3mm]{ }
       child{node [circle,draw, thick, minimum width=3mm]{ }}
       child{node [circle,draw, thick, minimum width=3mm]{ }}
      }
      child {node [black, fill=blue!50,circle,draw, thick, minimum width=3mm]{ }
       child{node [circle,draw, thick, minimum width=3mm]{ }}
       child{node [circle,draw, thick, minimum width=3mm]{ }}
       }
      child {node [black, fill=blue!50,circle,draw, thick, minimum width=3mm]{ }}
    }
    child {node[black, fill=blue!50,circle,draw, thick, minimum width=3mm]{ }
    child {node [black, fill=blue!50,circle,draw, thick, minimum width=3mm]{ }}
      child {node [black, fill=blue!50,circle,draw, thick, minimum width=3mm]{ }}
      child {node [black, fill=blue!50,circle,draw, thick, minimum width=3mm]{ }}
    };
\end{tikzpicture} \\
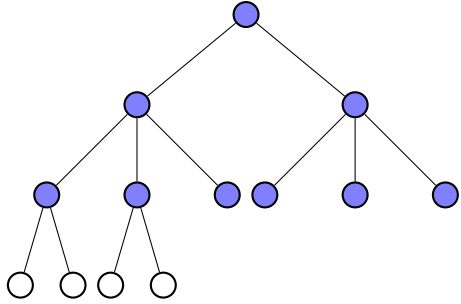
\footnotesize{Figure 4.1.1. Blue vertices represent a $(q-s)_e$ tree. Adding $2s$ leaves raises the total number of leaves to $q$.}
\end{center}
We can compute the number of $q_o$ trees by taking the number of possible $(q-s)_e$ trees and subtracting cases where $p<s$. Since the $p$ value that $s$ is being compared to represents the number of vertices on the top level of a $q_o$ tree, $p$ is always a multiple of 3. This means that the greatest $p$ value that $s=3j-2,3j-1,3j$ are all going to be greater than is the same. Thus the same number of terms are going to be subtracted for every triplet of $s$ values in that form. We can then write the number of cases of $t_n(p,(q-s)_e)$ that need to be subtracted as $\lfloor \frac{s-1}{3} \rfloor$ because plugging in the values $3j-2,3j-1,3j$ all result in the same number.

\paragraph{Theorem 4.1.2} $t_{2,3}(3s,q_e)=(q-2s)_o-\displaystyle\sum\limits_{p=1}^{\lfloor \frac{s-1}{2} \rfloor} t_{2,3}(p,(q-2s)_o)$
\paragraph{Proof:} Similar to our method of building a $q_o$ tree in Theorem 4.1.1, we consider constructing a $q_e$ by starting out with a $(q-2s)_o$ tree with $p$ leaves on its highest level, where $p \ge s$. We start with such a tree because adding three children onto $s$ of the top-level leaves causes a net increase of $2s$ leaves, which would give us a total of $q$ leaves. \\
\begin{center}
\begin{tikzpicture}[level distance=1.2cm,
  level 1/.style={sibling distance=3.0cm},
  level 2/.style={sibling distance=1.2cm},
  level 3/.style={sibling distance=0.7cm},
  level 4/.style={sibling distance=0.7cm}]
  \node [black, fill=blue!50,circle,draw, thick, minimum width=2mm](z){  }
    child {node[black, fill=blue!50,circle,draw, thick, minimum width=2mm]{ }
      child {
      node [black, fill=blue!50,circle,draw, thick, minimum width=2mm]{ }
       child{node [black, fill=blue!50,circle,draw, thick, minimum width=2mm]{ }
       child{node [circle,draw, thick, minimum width=2mm]{ }}
       child{node [circle,draw, thick, minimum width=2mm]{ }}
       child{node [circle,draw, thick, minimum width=2mm]{ }}
       }
       child{node [black, fill=blue!50,circle,draw, thick, minimum width=2mm]{ }}
      }
      child {node [black, fill=blue!50,circle,draw, thick, minimum width=2mm]{ }
       child{node [black, fill=blue!50,circle,draw, thick, minimum width=2mm]{ }}
       child{node [black, fill=blue!50,circle,draw, thick, minimum width=2mm]{ }}
       }
      child {node [black, fill=blue!50,circle,draw, thick, minimum width=2mm]{ }}
    }
    child {node[black, fill=blue!50,circle,draw, thick, minimum width=2mm]{ }
    child {node [black, fill=blue!50,circle,draw, thick, minimum width=2mm]{ }}
      child {node [black, fill=blue!50,circle,draw, thick, minimum width=2mm]{ }}
      child {node [black, fill=blue!50,circle,draw, thick, minimum width=2mm]{ }}
    };
\end{tikzpicture} \\
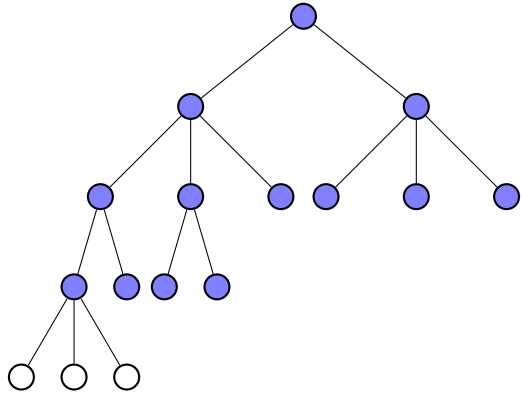
\footnotesize{Figure 4.1.2. Blue vertices represent a $(q-2s)_e$ tree. This time, three children are given to $s$ leaves.}
\end{center}
We can thus compute the number of $q_e$ trees by starting with the total number of $(q-2s)_o$ trees and subtracting the cases where $p<s$. In this computation, $s$ is being compared to $p$, the number of leaves on the bottom level of a $q_e$ tree, which means that $p$ has to be even. Then by the same logic in Theorem 4.1.1, when $s=2j-1,2j$, the number of cases of $p$ that need to be subtracted is same. We then find an expression that yields equal results when $2j-1$ and $2j$ are plugged in, which is $\lfloor \frac{s-1}{2} \rfloor$.
\subsubsection{The Recursive Formula.} 
\end{spacing}
Using Theorems 4.1.1 and 4.1.2, we can express the values of $q_o$ and $q_e$ as previous $(q-i)_o$ and $(q-i)_e$ terms and observe the patterns noted in these theorems. For example, the representation of $q_o$ is shown below.
\newline
\begin{spacing}{1.5}
\begin{center}
\begin{tabular}{l}
$q_o=$\\
$(q-1)_e+$\\
$(q-2)_e+$\\
$(q-3)_e+$\\
$(q-4)_e-(q-6)_o+$\\
$(q-5)_e-(q-7)_o+$\\
$(q-6)_e-(q-8)_o+$\\
$(q-7)_e-(q-9)_o-(q-11)_o+$\\
$(q-8)_e-(q-10)_o-(q-12)_o+$\\
$(q-9)_e-(q-11)_o-(q-13)_o+$\\
$(q-10)_e-(q-12)_o-(q-14)_o-[(q-16)_o-(q-17)_e]+$\\
$(q-11)_e-(q-13)_o-(q-15)_o-[(q-17)_o-(q-18)_e]+$\\
$(q-12)_e-(q-14)_o-(q-16)_o-[(q-18)_o-(q-19)_e]+$\\
... \\
\end{tabular}
\end{center}
\end{spacing}
\begin{spacing}{2.0}
We again use the technique of computing bounds up to a certain term $t_{2,3}(ni,q)$, where $n=2$ when the tree has an odd highest level and $n=3$ when the tree has an even highest level, for a given integer, $i$. We can express $q_o$ as several summations:
\begin{align*}
\displaystyle\sum\limits_{k=1}^i (q-k)_e &- \displaystyle\sum\limits_{t=0}^{i-4} (q-6-t)_o -\displaystyle\sum\limits_{t=0}^{i-7} (q-11-t)_o ...\\ &+ \displaystyle\sum\limits_{v=0}^{i-10} (q-17-v)_e + \displaystyle\sum\limits_{v=0}^{i-16} (q-28-v)_e...
\end{align*}

Putting these together yields
\newline $q_o={\displaystyle\sum\limits_{k=1}^{i} (q-k)_e}
-{\displaystyle\sum\limits_{r=1}^{\lfloor \frac{i-1}{3} \rfloor}
{\textstyle\sum\limits_{t=o}^{n-3r-1} (q-5r-1-t)_o}}
+{\displaystyle\sum\limits_{w=1}^{\lfloor \frac{i-4}{6} \rfloor}
{\textstyle\sum\limits_{v=o}^{n-6w-4} (q-11w-6-v)_e}}$.\\
\newline In a similar method, it can be reached that
\newline
$q_e=\displaystyle\sum\limits_{k=1}^i h(q-2k)_o-\displaystyle\sum\limits_{m=1}^{\lfloor \frac{i-1}{2} \rfloor} \textstyle\sum\limits_{f=0}^{i-2m-1} h(q-5m-2-2f)_e+\displaystyle\sum\limits_{b=1}^{\lfloor \frac{i-7}{2} \rfloor} \textstyle\sum\limits_{c=0}^{i-2b-7} h(q-5b-19-2c)_o$. \\ 
\newline The characteristic root method for finding an explicit formula for a recursion is based the fact that all terms are of the same sequence. This means that we cannot find numerical bounds on BT trees if we have $q_o$ written in terms of a mix of $(q-i)_o$ and $(q-i)_e$ terms, and vice versa. We must find some way to express $q_o$ in strictly $(q-i)_o$ terms and $q_e$ in strictly $(q-i)_e$ terms. Theorems 4.2 and 4.3 that follow aim to take a step in reaching that goal.

\paragraph{Theorem 4.2} $(q+1)_o \le q_o+q_e$ 
\subparagraph{Proof:} If we want to plug $q+1$ into the current formula for computing $q_e$, we're essentially keeping the lower bound but shifting the index of everything else by 1. Thus we need to modify the current formula by adding 1 to all the limits because $(q-i)$ is the smallest term that we want to include in our representation. Then we get:\\
$(q+1)_o=
    {\displaystyle\sum\limits_{k=1}^{n+1} (q+1-k)_e}-
    {\displaystyle\sum\limits_{r=1}^{\lfloor \frac{n}{3} \rfloor} \textstyle\sum\limits_{t=o}^{n-3r} (q-5r-t)_o}+
    {\displaystyle\sum\limits_{w=1}^{\lfloor \frac{n-3}{6} \rfloor} \textstyle\sum\limits_{v=o}^{n-6w-3} (q-11w-5-v)_e}$.
   
We then separate the proof into three cases of $q$.\\
\newline
\textbf{Case 1:} $q \equiv 1,2,4,5$ (mod 6). Since ${\lfloor \frac{n}{3} \rfloor} = {\lfloor \frac{n-1}{3} \rfloor}$ and ${\lfloor \frac{n-3}{6} \rfloor} = {\lfloor \frac{n-4}{6} \rfloor}$, substitution yields
\begin{center}
$(q+1)_o-q_o=q_e-\left({\displaystyle\sum\limits_{r=1}^{\lfloor \frac{n}{3} \rfloor} (q-5r)_o}-{\displaystyle\sum\limits_{w=1}^{\lfloor \frac{n-3}{6} \rfloor} (q-11w-5)_o}\right)$.
\end{center}
It is clear that
\begin{center}
${\displaystyle\sum\limits_{r=1}^{\lfloor \frac{n}{3} \rfloor} (q-5r)_o} \ge {\displaystyle\sum\limits_{r=1}^{\lfloor \frac{n-3}{6} \rfloor} (q-5r)_o} \ge {\displaystyle\sum\limits_{w=1}^{\lfloor \frac{n-3}{6} \rfloor} (q-11w-5)_o}$. \end{center}
From this, we can see that
\begin{center}
${\displaystyle\sum\limits_{r=1}^{\lfloor \frac{n}{3} \rfloor} (q-5r)_o} -{\displaystyle\sum\limits_{w=1}^{\lfloor \frac{n-3}{6} \rfloor} (q-11w-5)_o} \ge 0$.
\end{center}
Thus, $(q+1)_o-q_o \le q_e$. \\
 \textbf{Case 2:}    $q \equiv 0$ (mod 6). Since ${\lfloor \frac{n}{3} \rfloor} \neq {\lfloor \frac{n-1}{3} \rfloor}$, then ${\lfloor \frac{n}{3} \rfloor}={\lfloor \frac{n-1}{3} \rfloor}+1$. Since ${\lfloor \frac{n-3}{6} \rfloor} = {\lfloor \frac{n-4}{6} \rfloor}$ is also true, the $(q+1)_o$ term in this case is greater then the $(q+1)_o$ term in Case 1. Thus the claim holds true. \\
\textbf{Case 3:} $q \equiv 3$ (mod 6). Now that ${\lfloor \frac{n-3}{6} \rfloor} = {\lfloor \frac{n-4}{6} \rfloor}+1$ and ${\lfloor \frac{n}{3} \rfloor}={\lfloor \frac{n-1}{3} \rfloor}+1$, the $(q+1)_o$ term is greater than in Case 2, and thus the claim holds true as well. \\
\newline This theorem concludes $(q+1)_o-(q_o) \le q_e$. When there are $(q-i)_e$ terms in the representation of $q_o$, this inequality could potentially be used to substitute all the $(q-i)_e$ for $(q-i)_o$ terms to get a characteristic equation that would determine a lower bound.
\paragraph{Theorem 4.3} $(q+2)_e \le q_o+q_e$

\subparagraph{Proof:} If we want to plug $q+2$ into the current formula for computing $q_e$, then we need to modify the current formula by adding 2 to all the limits because the smallest term we want the answer to be expressed in terms of is $q-i$. The result is\\
$(q+2)_e=
    {\displaystyle\sum\limits_{k=1}^{\lfloor \frac{n+2}{2} \rfloor} (q+2-2k)_o}-
    {\displaystyle\sum\limits_{m=1}^{\lfloor \frac{n+1}{2} \rfloor} {\textstyle\sum\limits_{f=0}^{n-2m+1} (q-5m-2f)_e}}+
    {\displaystyle\sum\limits_{b=1}^{\lfloor \frac{n-5}{2} \rfloor} {\textstyle\sum\limits_{c=0}^{n-2b-5} (q-5b-2c-17)_o}}$ \\
    \newline
Now we do the subtraction and get \\
$(q+2)_e-q_e =
    \left(q+2-2\lfloor \frac{n+2}{2} \rfloor \right)_o-
    \displaystyle\sum\limits_{f=0}^{1} (q-5\left(\lfloor \frac{n+1}{2} \rfloor \right)-2f)_e+
    \displaystyle\sum\limits_{c=0}^{1} (q-5\left(\lfloor \frac{n-5}{2} \rfloor \right)-2c-17)_o$.
    \newline \newline
Using the logic similar to that of Theorem 4.2, it is clear that \\
$\displaystyle\sum\limits_{f=0}^{1} (q-5\left(\lfloor \frac{n+1}{2} \rfloor \right)-2f)_e-
    \displaystyle\sum\limits_{c=0}^{1} (q-5\left(\lfloor \frac{n-5}{2} \rfloor \right)-2c-17)_o \ge 0$. \\
    \newline
    Additionally, we also can note that $q_o \ge \left(q+2-2\lfloor \frac{n+2}{2} \rfloor \right)_o$. Setting
    \begin{center}
     $\alpha=\displaystyle\sum\limits_{f=0}^{1} (q-5\left(\lfloor \frac{n+1}{2} \rfloor \right)-2f)_e-
    \displaystyle\sum\limits_{c=0}^{1} (q-5\left(\lfloor \frac{n-5}{2} \rfloor \right)-2c-17)_o$
    \end{center}
So $q_o \ge q_o-\alpha \ge \left(q+2-2\lfloor \frac{n+2}{2} \rfloor \right)_o-\alpha$.
     Thus $q_o \ge (q+2)_e-q_e$ and the result follows. \\
     This theorem accomplishes something similar to Theorem 4.2. It provides an inequality that can be used to substitute $(q-i)_o$ terms for $(q-i)_e$ terms and potentially reach a lower bound, since what we're substituting in is smaller than the original term. \\
\section{Future Research}
Through our research we devise a recursive formula that can be used to represent BT trees. In order to generate numeric bounds, we must be able to cancel out all the $(q-i)_o$ terms in the representation of $q_e$ and vice versa. Theorems 4.2 and 4.3 that we provide are for this purpose. However, we believe that more accurate results can be achieved. Additionally, we plan to study more BT trees such as $2,2,3,3$ trees, $3, 2$ trees and other trees whose sequence have different combinations of 2s and 3s. For $3,2$ trees, we can use a similar approach to that of a $2,3$ tree. However, for other trees (such as $2,2,3,3$) we will have to consider $k$ cases, where $k$ is the period of the sequence. We conjecture, and hope to prove, that if the ratio between 2s and 3s in a sequence are equivalent for various BT trees, then the trees will have similar or identical bounds.

\section{Conclusion}
 The first portion of our results gives an algorithm for finding the upper and lower bounds of the number of $q$-length $n$-ary Huffman sequences with a specified accuracy. The algorithm simplifies the method of reaching an approximation of $h_n(q)$ that matches the currently most accurate results. The method we prove leads to our research in the second portion of our results. The second part opens a completely new door in the study of Huffman sequences by enumerating trees with vertices that may have 2 or 3 children, depending on the level that they are located on. We start by splitting up the Binary-Ternary trees into the cases based on whether the highest level is even or odd. We then began to show many patterns such a tree holds that are similar to patterns in $n$-ary Huffman trees. To the best of our knowledge, this is the first paper that studies BT trees. Our results have significant applications in computer data compression and in-depth genome analysis.

\section{Acknowledgement}
This is part of a research project done by three high school students (Rao, Liu, and Feng) in the summer of 2012 under the supervision of Dr. Jian Shen at Texas State University.
Rao, Liu, and Feng thank Texas State Math Camp for providing this research opportunity. Rao wants to thank Professor H. Prodinger for providing reference [5].
\end{spacing}

\end{spacing}
\end{document}